\documentclass[twocolumn,floatfix,superscriptaddress,showpacs,showkeys]{revtex4}
\usepackage{amsmath}
\usepackage{graphicx}
\usepackage{dcolumn}
\usepackage{bm}
\usepackage{booktabs}
\usepackage{amssymb}
\usepackage{multirow}
\usepackage{makecell}
\usepackage{textcomp}
\usepackage{subfigure}
\usepackage{phonetic}
\usepackage{extarrows}
\usepackage{color}
\usepackage{float}
\usepackage[colorlinks,citecolor=blue,linkcolor=blue]{hyperref}

\begin{document}


\title{Controllable double quantum state transfers by one topological channel in a frequency-modulated optomechanical array}

\author{Lu Qi}
\affiliation{Department of Physics, Harbin Institute of Technology, Harbin, Heilongjiang 150001, China}
\author{Shutian Liu}
\email{stliu@hit.edu.cn}
\affiliation{Department of Physics, Harbin Institute of Technology, Harbin, Heilongjiang 150001, China}
\author{Shou Zhang}
\email{szhang@ybu.edu.cn}
\affiliation{Department of Physics, Harbin Institute of Technology, Harbin, Heilongjiang 150001, China}
\affiliation{Department of Physics, College of Science, Yanbian University, Yanji, Jilin 133002, China}
\author{Hong-Fu Wang}
\email{hfwang@ybu.edu.cn}
\affiliation{Department of Physics, College of Science, Yanbian University, Yanji, Jilin 133002, China}


\date{\today}

\begin{abstract}
We propose a scheme to achieve the quantum state transfer via the topological protected edge channel based on a one dimensional frequency-modulated optomechanical array. We find that the optomechanical array can be mapped into a Su-Schrieffer-Heeger model after eliminating the counter rotating wave terms via frequency modulations. By dint of the edge channel of the Su-Schrieffer-Heeger model, we show that the quantum state transfer between the photonic left edge state and the photonic right edge state can be achieved with a high fidelity. Specially, our scheme can also achieve another phononic quantum state transfer based on the same channel via controlling the next-nearest-neighboring interactions between the cavity fields, which is different from the previous investigations only achieving one kind of quantum state transfer. Our scheme provides a novel path to switch two different kinds of quantum state transfers in a controllable way. 
\end{abstract}

\maketitle


\section{\label{sec.1}Introduction}
Topological insulator~\cite{hasan2010colloquium,qi2011topological,chiu2016classification,bansil2016colloquium} is characterized by the conducting edge states and the insulating bulk states. These conducting edge states are immune to the local disorders and perturbations since they are protected by the nonlocal topological invariant~\cite{matsuura2010momentum,wray2011topological,nussinov2009sufficient,shapourian2017many}. The topological insulator has many potential applications in quantum information processing and quantum computing. For example, we can achieve the robust quantum state transfer via the edge channel of the topological insulator~\cite{dlaska2017robust,bonderson2011topological,leijnse2011quantum}, in which the process of state transfer is robust to the local disorders and perturbations since the edge channel is protected by the energy gap. Also, we can implement the topological quantum computing~\cite{freedman2003topological,sau2010generic} via the non-Abelian anyons~\cite{nayak2008non} and Majorana zero modes~\cite{sarma2015majorana}, etc. The Su-Schrieffer-Heeger (SSH) model, as the simplest one dimensional (1D) tight-binding topological insulator model, has attracted a great attention due to the appearance of the abundant physics, such as the topological phase transition~\cite{fradkin1983phase,marchand2010sharp,li2014topological}, the edge state and topological invariant~\cite{fu1988bound,dangel2018topological,yao2018non}, $\mathcal{PT}$- symmetry effect~\cite{zhu2014pt,klett2018topological,klett2017relation,ke2017topological}, the photonic mapping of SSH model~\cite{longhi2019probing,tan2014photonic,mei2015simulation}, non-hermitian bulk-boundary correspondence~\cite{lieu2018topological,kunst2018biorthogonal,yao2018edge}, etc. Specially, the quantum state transfer between the left edge and the right edge states by the topological zero mode of a photonic topological SSH model has also been investigated~\cite{mei2018robust}. 

In the previous works mentioned above, only one kind of quantum state transfer can be achieved via the topological channel. In this paper, we propose a scheme to achieve two different kinds of quantum state transfers only by one topological channel based on a frequency-modulated optomechanical array. Note that the optomechanical system with frequency modulation has been intensively investigated in experiment~\cite{singh2014optomechanical}. We show that the two different kinds of quantum state transfers can be achieved via controlling the next-nearest-neighboring (NNN) hopping. Specially, the two different kinds of quantum state transfers can be switched to each other in a controllable way, which is expected to greatly improve the efficiency of quantum information processing.

\section{\label{sec.2}System and Hamiltonian}
\begin{figure}
	\centering
	\includegraphics[width=0.9\linewidth]{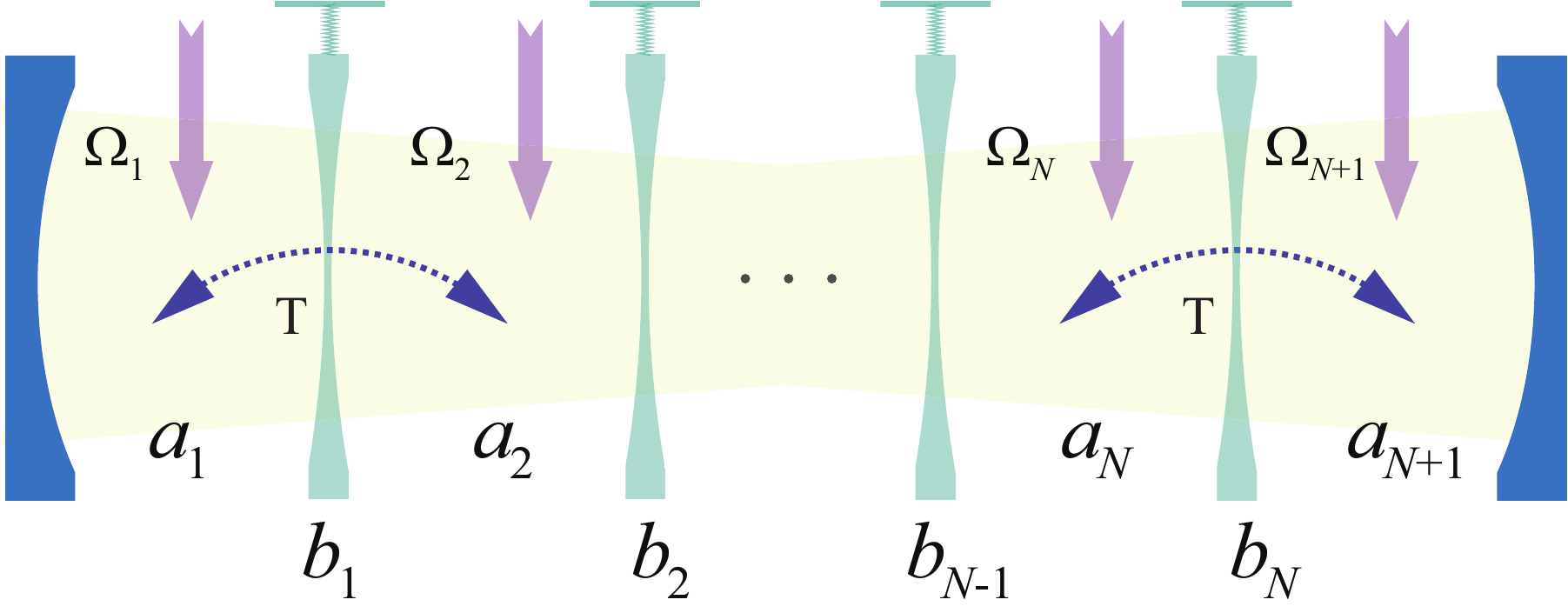}\\
	\caption{Schematic of the 1D optomechanical array. The multi-resonator optomechanical system contains $N+1$ cavity modes and $N$ resonators. Each cavity mode is driven by a laser field and the coupling between resonator $b_{n}$ and cavity field $a_{n}$ ($a_{n+1}$) is $g_{n}$.}\label{fig1}
\end{figure}
\begin{figure}
	\centering
	\subfigure{\includegraphics[width=0.49\linewidth]{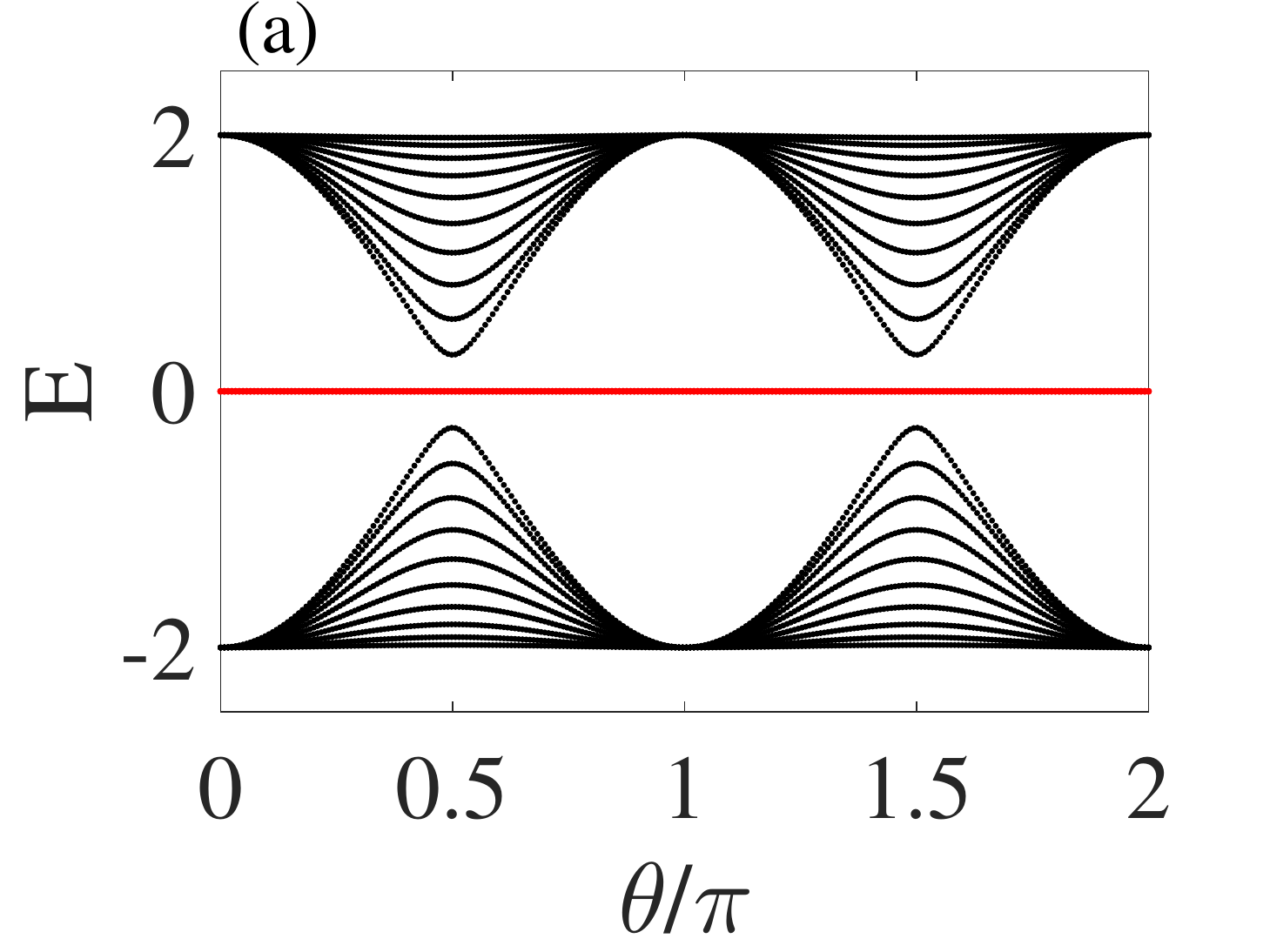}}
	\subfigure{\includegraphics[width=0.49\linewidth]{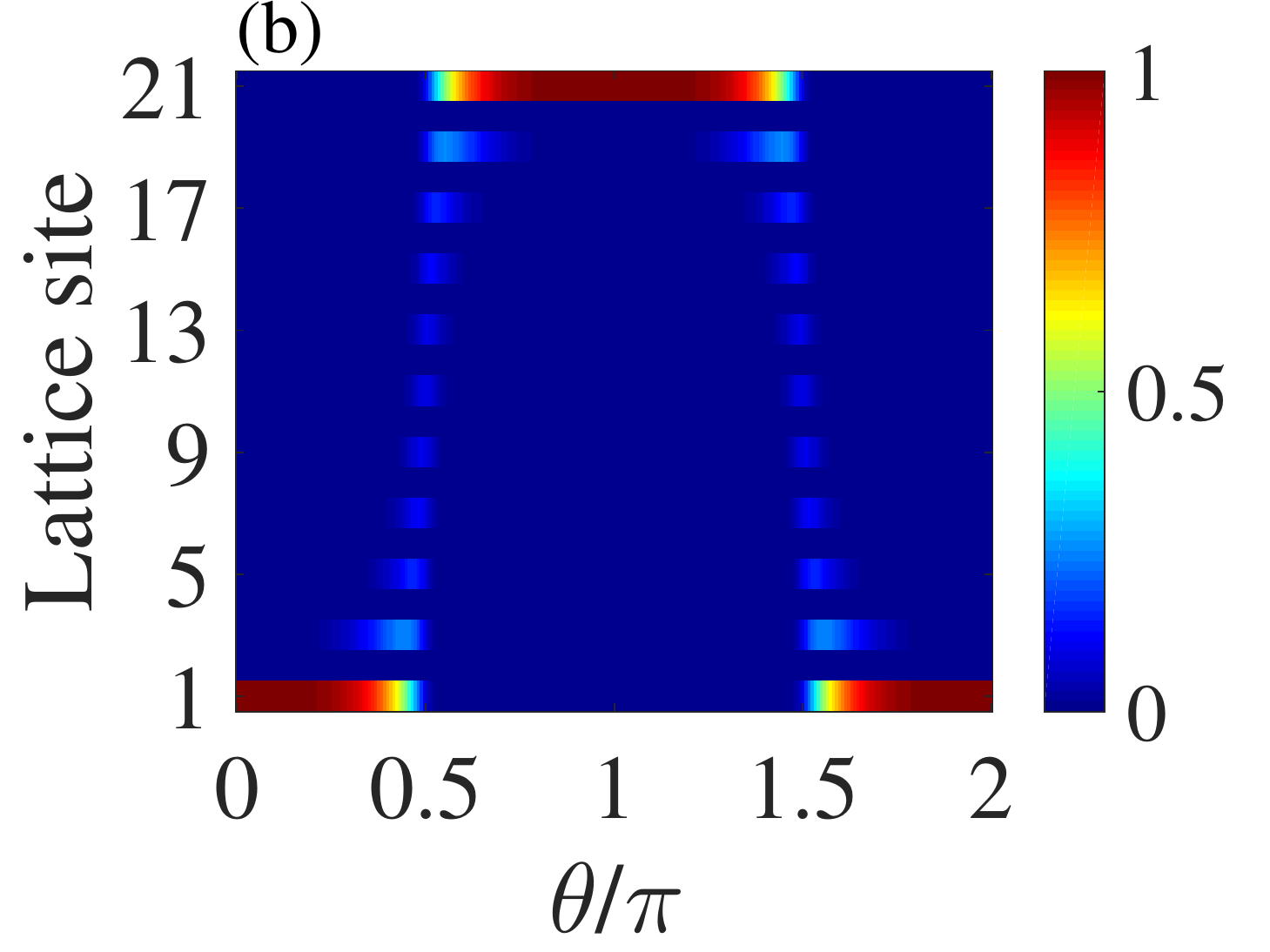}}
	\caption{Energy spectrum and the state distribution of the system. (a) Energy spectrum of the system. There is a zero energy mode in the gap represented by the red line. (b) The distribution of the zero energy mode. The size of the system is $L=2N+1=21$.}\label{fig2}
\end{figure}
\begin{figure}
	\centering
	\subfigure{\includegraphics[width=0.49\linewidth]{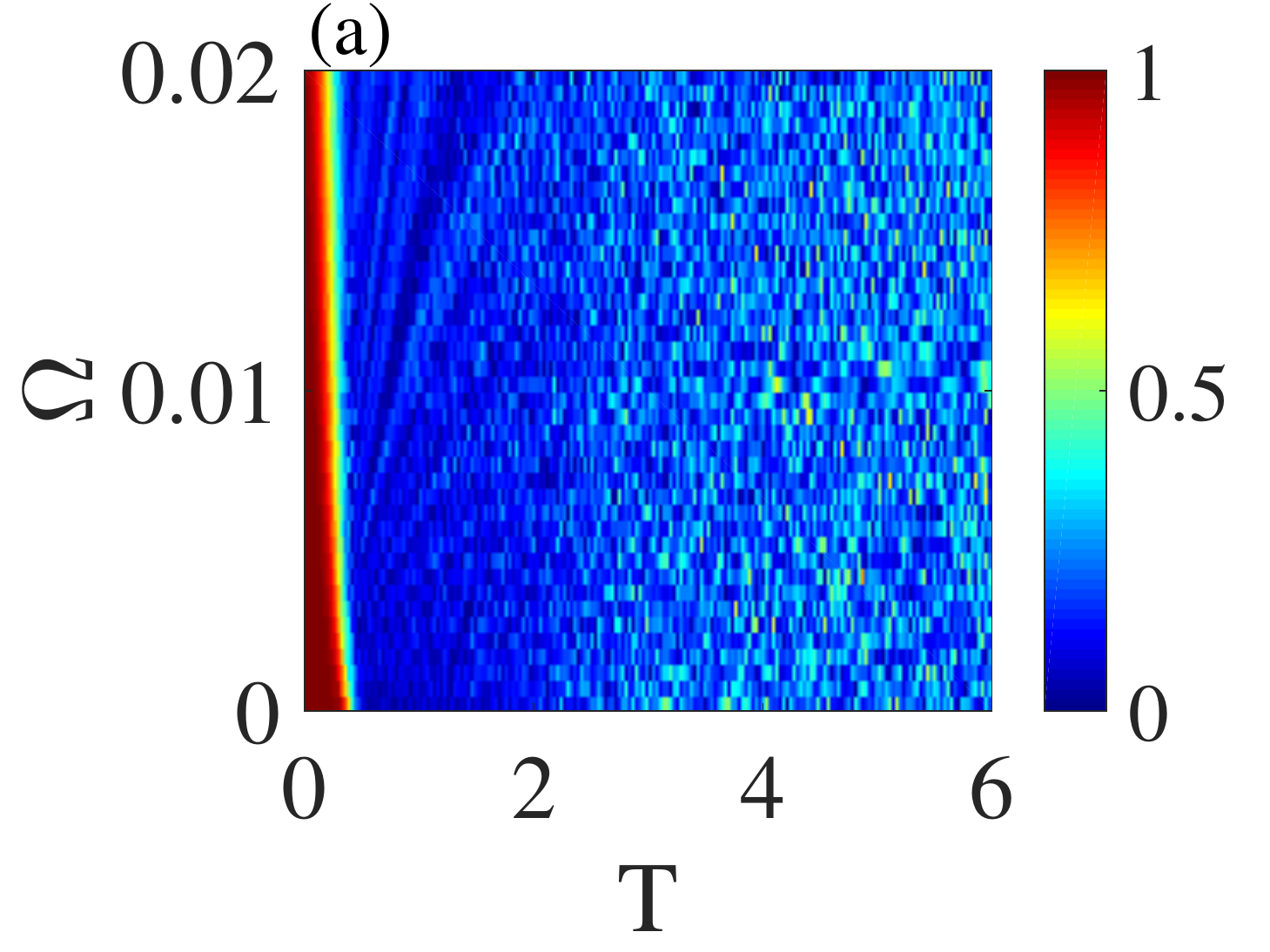}}
	\subfigure{\includegraphics[width=0.49\linewidth]{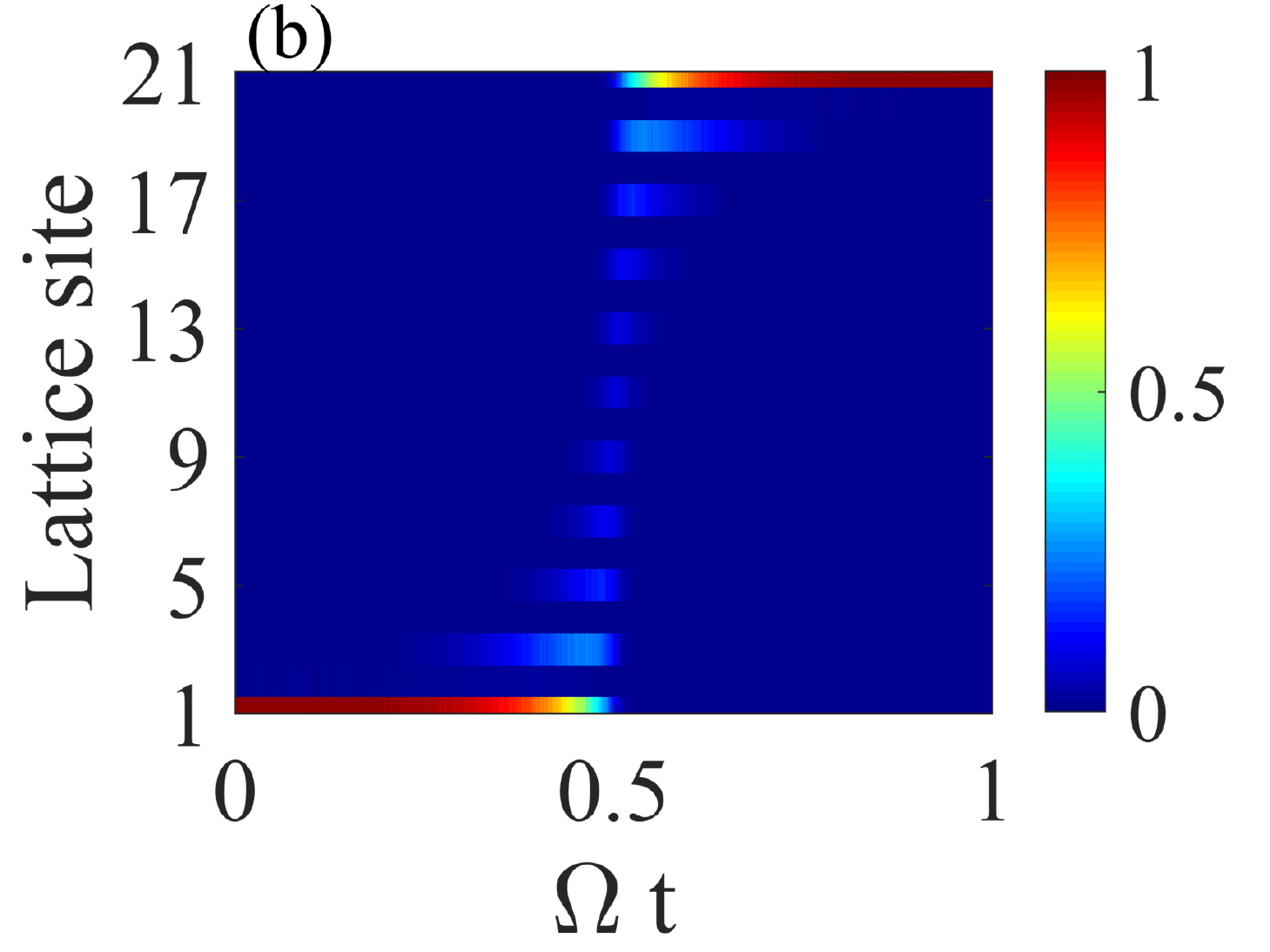}}
	\caption{The fidelity and the state transfer of the system. (a) The fidelity versus the varying rate $\Omega$ and the partial NNN hopping strength $T$. (b) The state transfer between $|L\rangle$ and $|R\rangle$ when $\Omega=0.001$ and $T=0.2$.}\label{fig3}
\end{figure} 

We consider a 1D optomechanical array containing $N+1$ cavity fields and $N$ resonators, in which each cavity field is driven by a laser with frequency $\omega_{d}$ and strength $\Omega_{n}$, as shown in Fig.~\ref{fig1}. In this array, the frequency of each cavity field and resonator can be modulated with the modulated frequency $\omega_{a,n}^{'}=\omega_{a,n}+\lambda_{n} \nu\cos(\nu t+\phi)$ and $\omega_{b,n}^{'}=\omega_{b,n}+\gamma_{n} \nu\cos(\nu t+\phi)$. The single-phonon optomechanical coupling strength between resonator $b_{n}$ and cavity field $a_{n}$ ($a_{n+1}$) is $g_{n}$. Moreover, the two adjacent cavity fields possess direct coupling with the coupling strength $T$. Then, the system can be described by the following Hamiltonian
\begin{eqnarray}\label{e01}
H&=&\sum_{n=1}^{N+1}\omega_{a,n}^{'}a_{n}^{\dag}a_{n}+\sum_{n=1}^{N} \omega_{b,n}^{'}b_{n}^{\dag}b_{n}\cr\cr
&&+\sum_{n=1}^{N+1}(\Omega_{n}a_{n}^{\dag}e^{-i\omega_{d}t}+\Omega_{n}^{\ast}a_{n}e^{i\omega_{d}t})\cr\cr
&&-\sum_{n=1}^{N}g_{n} (a_{n}^{\dag}a_{n}-a_{n+1}^{\dag}a_{n+1})(b_{n}^{\dag}+b_{n})\cr\cr
&&+\sum_{n=1}^{N}T(a_{n+1}^{\dag}a_{n}+a_{n}^{\dag}a_{n+1}),
\end{eqnarray}
where $a_{n}^{\dag}$ and $a_{n}$ ($b_{n}^{\dag}$ and $b_{n}$) are the creation and annihilation operators of the optical cavity field (mechanical resonator). The first two terms are the modulated free energy of the cavity fields and resonators, where $\lambda_{n}$ ($\gamma_{n}$), $\nu$, and $\phi$ are the modulated strength, frequency, and phase respectively. The third term describes the external driving of the cavity fields. The fourth term is the coupling between the cavity field and mechanical resonator via the radiation pressure. And the last term describes the direct tunneling between two adjacent cavity fields.

Under the condition of strong laser driving, we use the driving frequency $\omega_{d}$ to perform the rotating transformation and implement the standard linearization process via rewriting the operators as $a_{n}=\left \langle a_{n} \right \rangle+\delta a_{n}=\alpha_{n}+\delta a_{n}$ ($b_{n}=\left \langle b_{n} \right \rangle+\delta b_{n}=\beta_{n}+\delta b_{n}$). After dropping the notation ``$\delta$'' for all the fluctuation operators $\delta a_{n}$ ($\delta b_{n}$), the Hamiltonian can be rewritten as
\begin{eqnarray}\label{e02}
H_{L}&=&\sum_{n=1}^{N+1} \left[\Delta_{a,n}^{'}+\lambda_{n} \nu\cos(\nu t+\phi)\right]a_{n}^{\dag}a_{n}\cr\cr
&&+\sum_{n=1}^{N} \left[\omega_{b,n}+\gamma_{n} \nu\cos(\nu t+\phi)\right]b_{n}^{\dag}b_{n}-g_{n} (\alpha_{n}^{\ast}a_{n}\cr\cr
&&+\alpha_{n}a_{n}^{\dag}-\alpha_{n+1}^{\ast}a_{n+1}-\alpha_{n+1}a_{n+1}^{\dag})(b_{n}^{\dag}+b_{n})\cr\cr
&&+T(a_{n+1}^{\dag}a_{n}+a_{n}^{\dag}a_{n+1}),
\end{eqnarray}
where $\Delta_{a,n}^{'}$ is effective detuning originating from optomechanical coupling with $\Delta_{a,1}^{'}=\Delta_{a,1}-g_{1}(\beta_{1}^{\ast}+\beta_{1})$, $\Delta_{a,N+1}^{'}=\Delta_{a,N+1}+g_{N}(\beta_{N}^{\ast}+\beta_{N})$, $\Delta_{a,n=2...N}^{'}=\Delta_{a,n}-g_{n-1}(\beta_{n-1}^{\ast}+\beta_{n-1})+g_{n}(\beta_{n}^{\ast}+\beta_{n})$, and $\Delta_{a,n}=\omega_{a,n}-\omega_{d}$ is the detunings between cavity fields and driving fields. Then we perform another rotating transformation defined by $V=\mathrm{exp}\bigg\{\sum_{n=1}^{N+1}-i\Delta_{a,n}^{'}ta_{n}^{\dag}a_{n}-i\lambda_{n}\sin(\nu t+\phi) a_{n}^{\dag}a_{n}+\sum_{n=1}^{N}-i\omega_{b,n}tb_{n}^{\dag}b_{n}-i\gamma_{n} \sin(\nu t+\phi) b_{n}^{\dag}b_{n}\bigg\}.$
After that, the Hamiltonian becomes
\begin{eqnarray}\label{e03}
H_{L}^{(1)}&=&\sum_{n} \left[-G_{n}a_{n}^{\dag}b_{n}e^{i[(\Delta_{a,n}^{'}-\omega_{b,n})t+(\lambda_{n}-\gamma_{n})\sin(\nu t+\phi) ]}\right.\cr\cr
&&\left.-G_{n}a_{n}^{\dag}b_{n}^{\dag}e^{i[(\Delta_{a,n}^{'}+\omega_{b,n})t+(\lambda_{n}+\gamma_{n})\sin(\nu t+\phi) ]}\right.\cr\cr
&&+\left.G_{n+1}a_{n+1}^{\dag}b_{n}e^{i[(\Delta_{a,n+1}^{'}-\omega_{b,n})t+(\lambda_{n+1}-\gamma_{n})\sin(\nu t+\phi) ]}\right.\cr\cr
&&\left.+G_{n+1}a_{n+1}^{\dag}b_{n}^{\dag}e^{i[(\Delta_{a,n+1}^{'}+\omega_{b,n})t+(\lambda_{n+1}+\gamma_{n})\sin(\nu t+\phi) ]}\right.\cr\cr
&&+\left.T a_{n+1}^{\dag}a_{n} e^{i(\lambda_{n+1}-\lambda_{n})\sin (\nu t+\phi)}\right]+\mathrm{H.c.},
\end{eqnarray}
where $G_{n}=g_{n}\alpha_{n}$ ($G_{n+1}=g_{n}\alpha_{n+1}$) is the effective optomechanical coupling. We choose the parameters of the system to satisfy $\Delta_{a,n}^{'}=\Delta_{a,n+1}^{'}=\omega_{b,n}$, $\lambda_{n+1}=\lambda_{n}=\gamma_{n}$, and  $\phi=0$. Then, the Hamiltonian in Eq.~(\ref{e03}) becomes
\begin{eqnarray}\label{e04}
H_{L}^{(2)}&=&\sum_{n} \left[-G_{n}a_{n}^{\dag}b_{n}-G_{n}a_{n}^{\dag}b_{n}^{\dag}e^{i[2\omega_{b,n}t+2\lambda_{n}\sin(\nu t) ]}\right.\cr\cr
&&\left.+G_{n+1}a_{n+1}^{\dag}b_{n}+G_{n+1}a_{n+1}^{\dag}b_{n}^{\dag}e^{i[2\omega_{b,n}t+2\lambda_{n}\sin(\nu t) ]}\right.\cr\cr
&&\left.+T a_{n+1}^{\dag}a_{n}\right]+\mathrm{H.c.}.
\end{eqnarray} 
After exploiting the Jacobi$-$Anger expansions $e^{i\kappa \sin\nu t}=\sum_{m=-\infty}^{\infty}J_{m}(\kappa)e^{im\nu t}$, the above Hamiltonian can be rewritten as
\begin{eqnarray}\label{e05}
H_{L}^{(3)}&=&\sum_{n} \left[-G_{n}a_{n}^{\dag}b_{n}+G_{n+1}a_{n+1}^{\dag}b_{n}\right.\cr\cr
&&\left.+\sum_{m_{1}=-\infty}^{\infty} -G_{n}J_{m_{1}}(\kappa)a_{n}^{\dag}b_{n}^{\dag}e^{i[2\omega_{b,n}+m_{1}\nu ]t}\right.\cr\cr
&&+\left.\sum_{m_{2}=-\infty}^{\infty}G_{n+1}J_{m_{2}}(\kappa)a_{n+1}^{\dag}b_{n}^{\dag}e^{i[2\omega_{b,n}+m_{2}\nu ]t}\right.\cr\cr
&&\left.+T a_{n+1}^{\dag}a_{n}\right]+\mathrm{H.c.},
\end{eqnarray} 
where $J_{m_{j}}$ is the $m_{j}$th order of Bessel function with $j=1,~2$ and $\kappa=2\lambda_{n}$ is the Bessel parameter. When $\nu\gg \omega_{b,n}$, we find that the system has the minimum detuning if $m_{1}=m_{2}=0$, leading that 
\begin{eqnarray}\label{e06}
H_{L}^{(4)}&=&\sum_{n} \left[-G_{n}a_{n}^{\dag}b_{n}+G_{n+1}a_{n+1}^{\dag}b_{n} -G_{n}J_{0}(\kappa)a_{n}^{\dag}b_{n}^{\dag}e^{2i\omega_{b,n}t}\right.\cr\cr
&&\left.+G_{n+1}J_{0}(\kappa)a_{n+1}^{\dag}b_{n}^{\dag}e^{2i\omega_{b,n} t}+T a_{n+1}^{\dag}a_{n}\right]+\mathrm{H.c.}.
\end{eqnarray} 
Apparently, we can safely remove the resonant counter rotating wave terms by choosing the appropriate value of $\kappa$ to satisfy $J_{0}(\kappa)=0$, which can effectively avoid the limit of the weak coupling condition. Then the system can be described by the following Hamiltonian, with
\begin{eqnarray}\label{e07}
H&=&\sum_{n} \left[-G_{n}a_{n}^{\dag}b_{n}+G_{n+1}a_{n+1}^{\dag}b_{n}+T a_{n+1}^{\dag}a_{n}\right]+\mathrm{H.c.}.  
\end{eqnarray} 

When $T=0$, the above Hamiltonian is equivalent to a standard SSH-type tight-binding Hamiltonian possessing only the nearest-neighboring couplings, in which the cavity field $a_{n}$ and the resonator $b_{n}$ are equivalent to two sites in a unit cell. In this way, the above optomechanical array can be regard as a SSH chain with the lattice size of $L=2N+1$. Specially, when $T\ne 0$, the term of $T a_{n+1}^{\dag}a_{n}$ represents the partial NNN hopping added into the tight-binding SSH model. The word of ``partial'' means that only the odd sites (cavity fields) have the couplings.   
Obviously, the above Hamiltonian is a periodically modulated SSH model if varying the effective optomechanical coupling strength $-G_{n}$ (intra-cell couplings) and $G_{n+1}$ (inter-cell couplings) in an alternative way for $T=0$, such as $-G_{n}=G_{0}(1-\cos\theta$) and $G_{n+1}=G_{0}(1+\cos\theta)$~\cite{qi2017simulating}. We stress that $G_{0}$ needs to satisfy $G_{0}\leqslant0.5\omega_{b,n}$ in the usual optomechanical system and we set  $G_{0}=1$ as the energy unit in the following. The energy spectrum and the distribution of the zero energy mode of the periodically modulated SSH model are shown in Fig.~\ref{fig2}. We find that the energy spectrum has a zero energy mode in the gap localized near the left edge when $\theta\in(0, \frac{\pi}{2}) \cup (\frac{3\pi}{2}, 2\pi)$, while for other regions of $\theta$, it is localized near the right edge. It means that we can achieve the state transfer between the left edge state and the right edge state by the zero energy mode via varying the periodic parameter $\theta$ from $0$ to $\pi$ over time. For example, we can choose the periodic parameter $\theta$ satisfying $\theta(t)=\Omega t$ with the varying rate $\Omega$ and the time $t$. Thus the Hamiltonian can be rewritten as 
\begin{eqnarray}\label{e09}
H_{\mathrm{SSH}}(t)&=&\sum_{n} \left[1-\cos(\Omega t)\right] a_{n}^{\dag}b_{n}+\left[1+\cos(\Omega t)\right]a_{n+1}^{\dag}b_{n}\cr\cr
&&+\mathrm{H.c.}.  
\end{eqnarray}  
If the initial state is prepared in the perfect photonic left edge state $|\Psi\rangle_{\mathrm{initial}} =|L \rangle=|1\rangle_{a_{1}}\otimes|0\rangle_{b_{1}}\otimes|0\rangle_{a_{2}}...\otimes|0\rangle_{b_{N}}\otimes|0\rangle_{a_{N+1}}=|1,~0,~0,...,0,~0\rangle$, the right edge state can be obtained via the evolution of the present time-dependent Hamiltonian with $i\frac{d}{dt}|\Psi\rangle_{\mathrm{initial}}=H_{\mathrm{SSH}}(t)|\Psi\rangle_{\mathrm{initial}}$~\cite{mei2018robust}. A question arises: can this kind of state transfer still be achieved when the NNN hopping $T$ is added into the system?
We simulate the fidelity between the final state and the perfect right edge state $|R \rangle=|0,~0,~0,...,0,~1\rangle$ versus the varying rate $\Omega$ and the partial NNN hoping strength $T$ numerically, as shown in Fig.~\ref{fig3}(a). The numerical results show that the state transfer between $|L\rangle$ and $|R\rangle$ can be achieved with a high fidelity when $\Omega<0.01$ and $T<0.4$. The condition of $\Omega<0.01$ originates from the adiabatic evolution condition, which means that the varying of $\theta$ should be as slow as possible. And the condition of $T<0.4$ is due to the topological protection of the energy gap. It implies that the state transfer channel (zero energy mode) does not be destroyed as long as the NNN hopping strength $T$ is not beyond the width of the gap. To further clarify it, we simulate the process of the state transfer numerically when the initial state is prepared in the state of $|L\rangle$ when $T=0.2$, as shown in figs.~\ref{fig3}(b). Obviously, when $T<0.4$, the sate transfer between $|L\rangle$ and $|R\rangle$ can still be achieved while it is destroyed completely when the NNN hoping strength $T$ is large enough.
\begin{figure}
	\centering
	\subfigure{\includegraphics[width=0.32\linewidth]{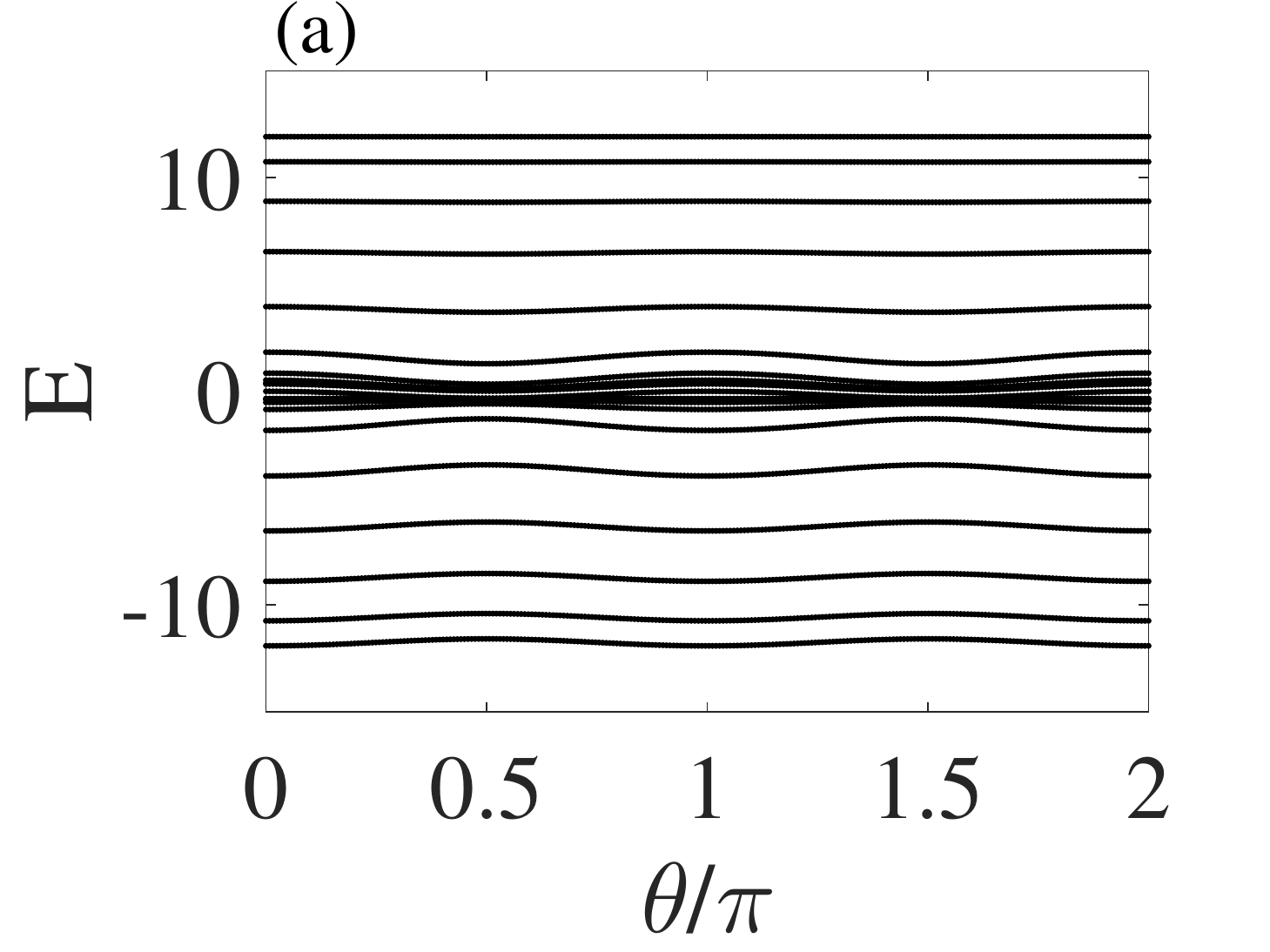}}
	\subfigure{\includegraphics[width=0.32\linewidth]{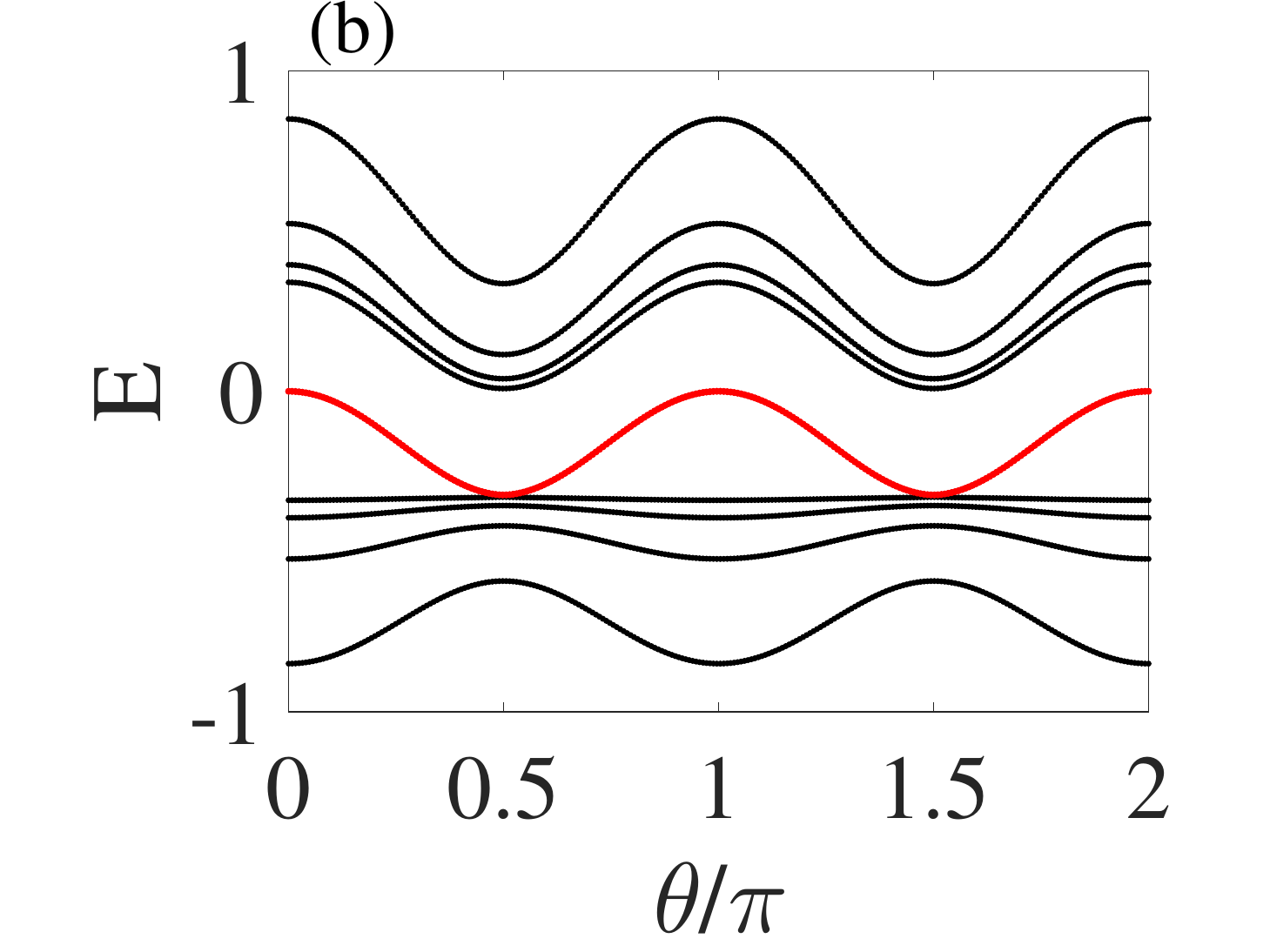}}
	\subfigure{\includegraphics[width=0.32\linewidth]{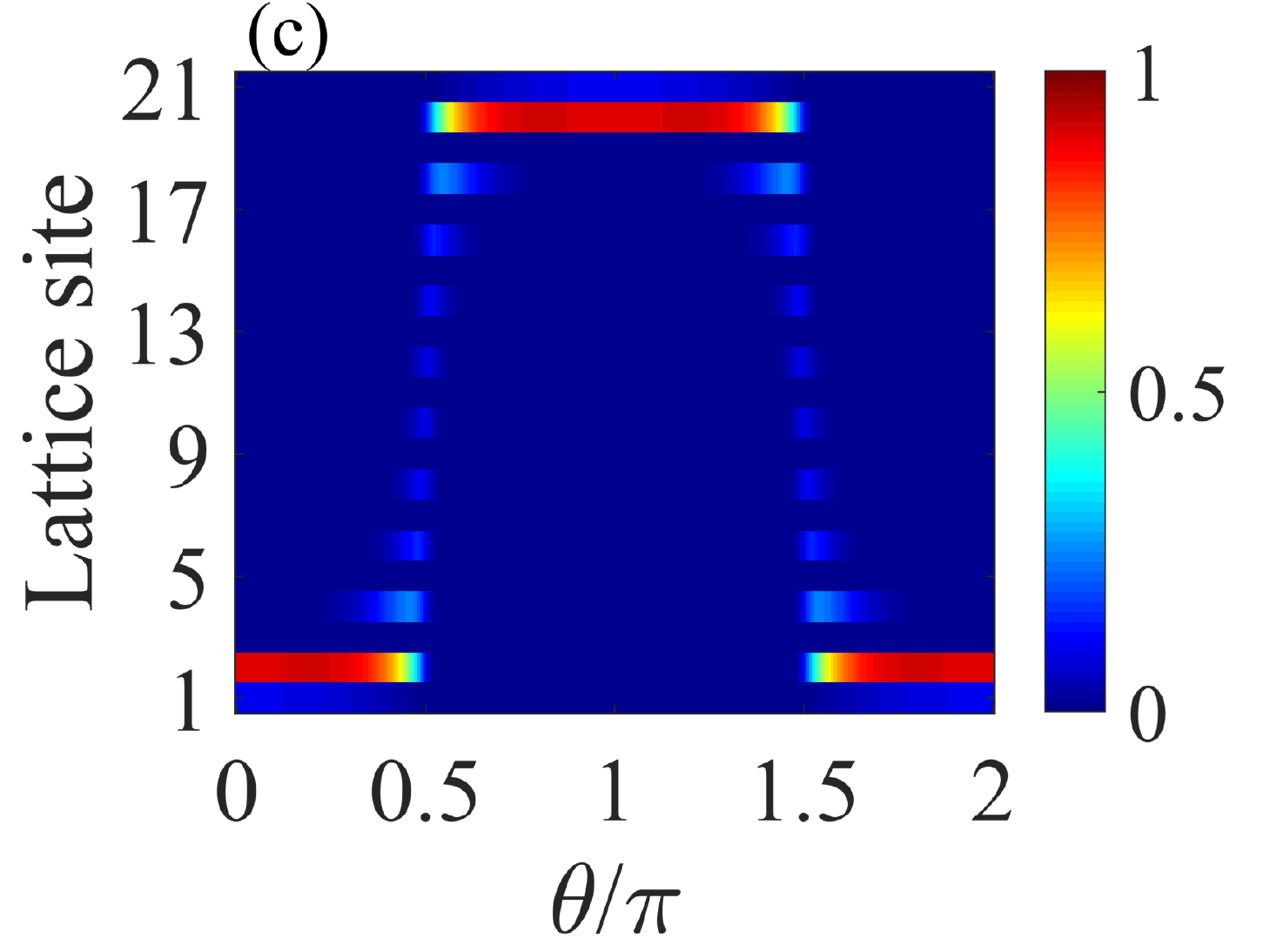}}
	\caption{The energy spectrum and the distribution of the gap state when $T=6$. (a) The energy spectrum of the system. (b) The detailed pattern in the energy gap. (c) The distribution of the gap state (red line in (b)).}\label{fig4}
\end{figure} 
\begin{figure}
	\centering
	\subfigure{\includegraphics[width=0.5\linewidth]{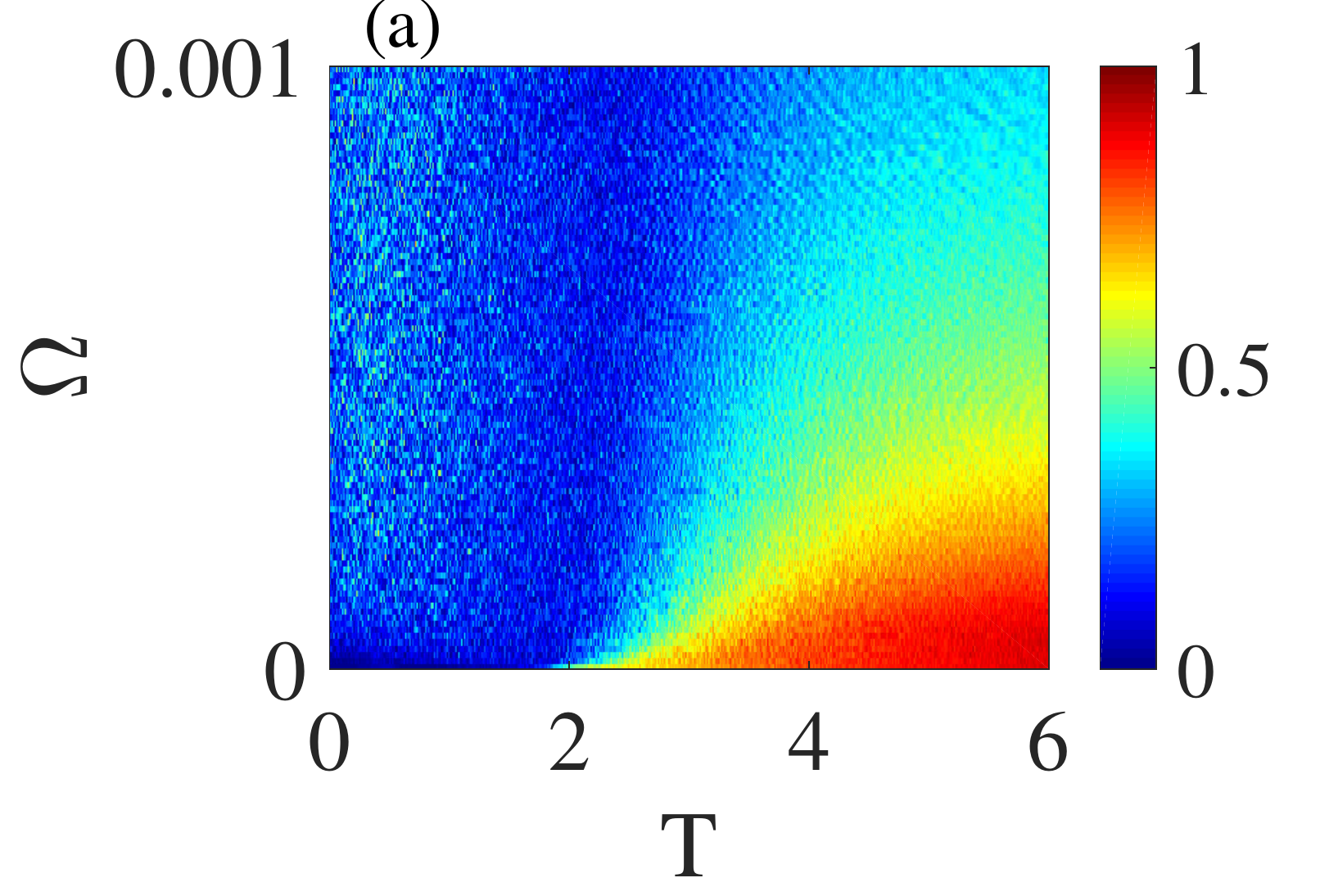}}
	\subfigure{\includegraphics[width=0.45\linewidth]{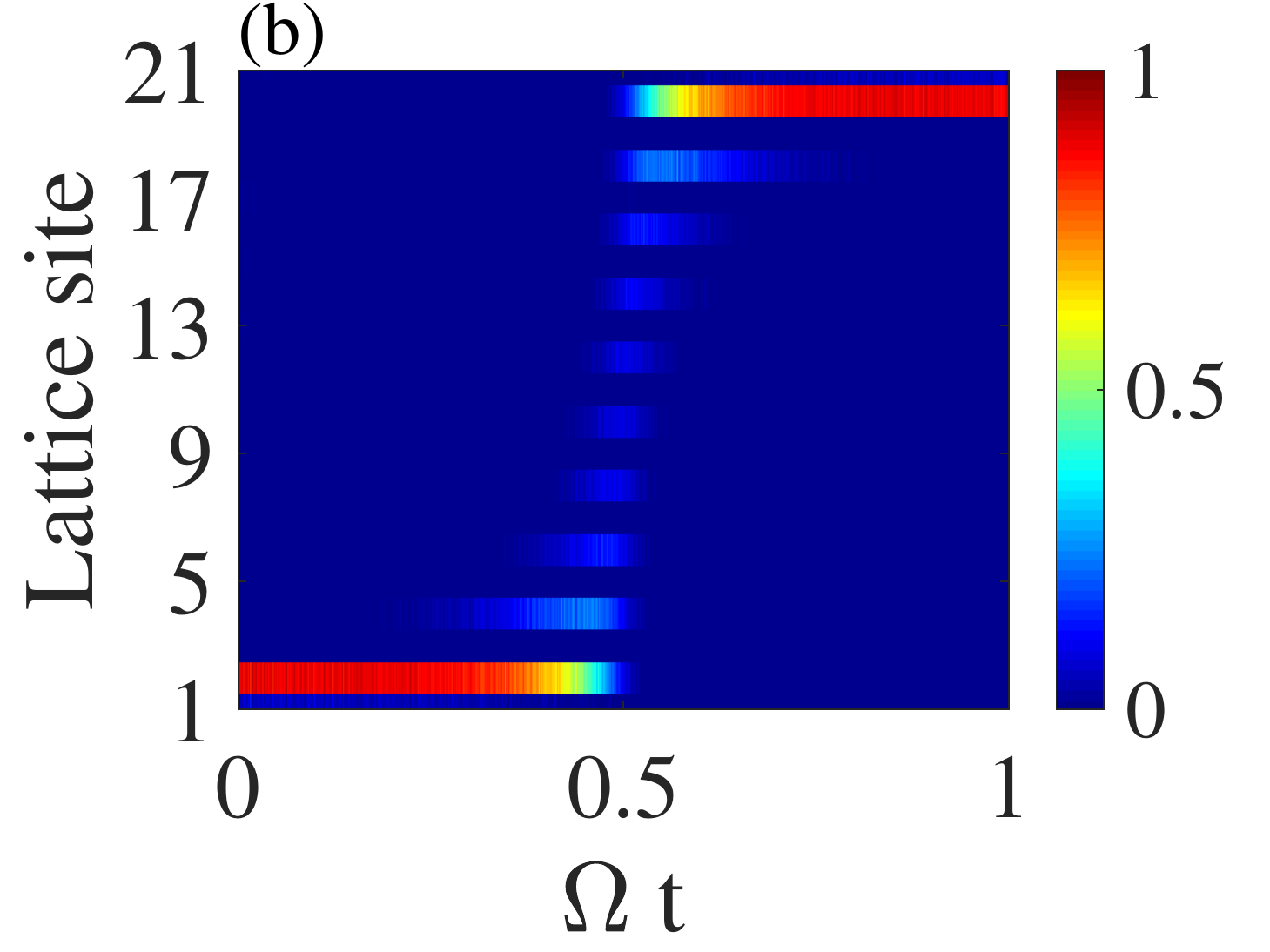}}
	\caption{The fidelity and the process of state transfer. (a) The fidelity between the final state and the state of $|R^{'}\rangle$ versus the varying rate $\Omega$ and the partial NNN hopping strength $T$. (b) The evolution of the initial state $|L^{'}\rangle$ over time when $T=6$ and $\Omega=0.00001$.}\label{fig5}
\end{figure} 
\begin{figure}
	\centering
	\subfigure{\includegraphics[width=0.49\linewidth]{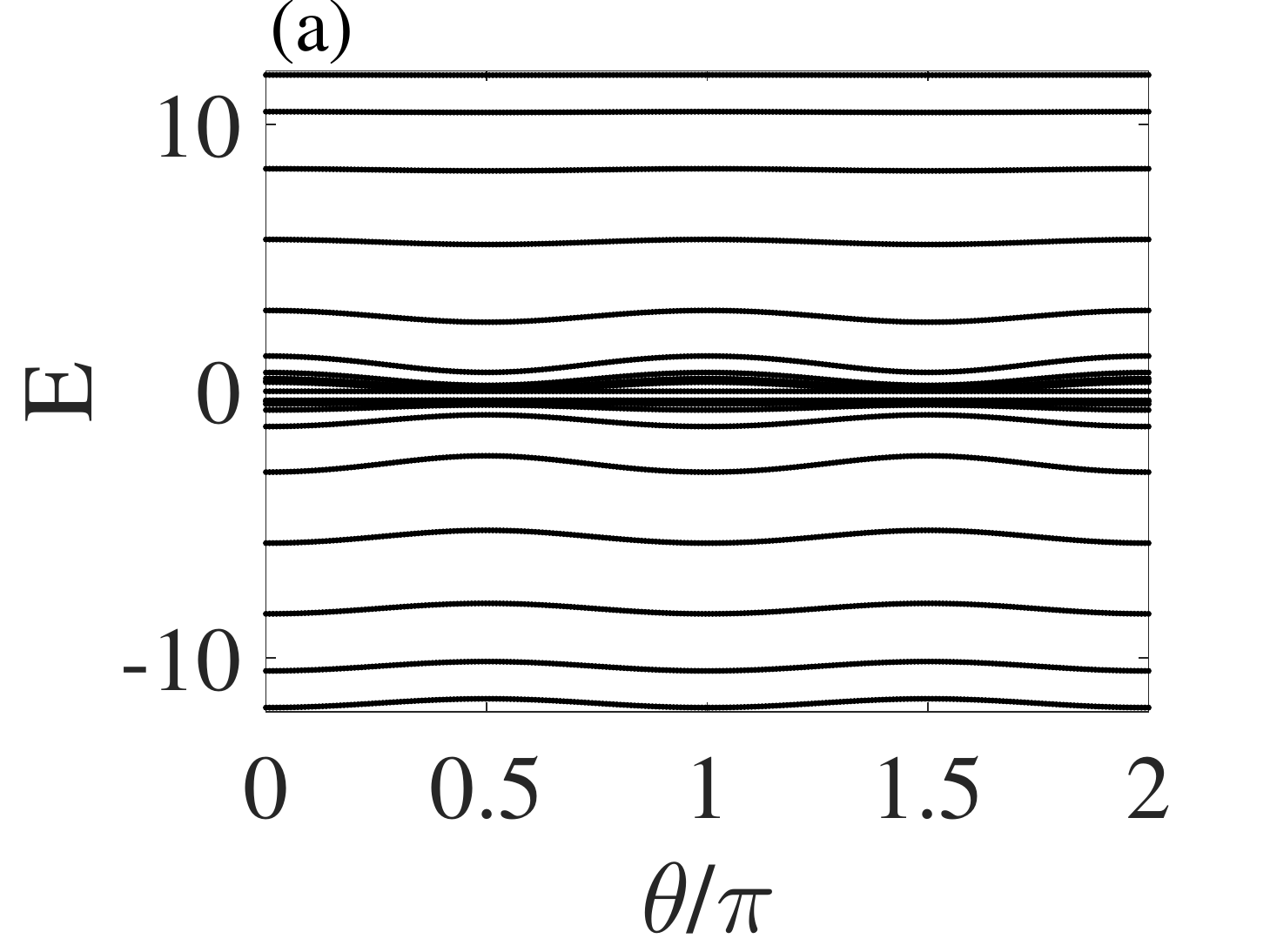}}
	\subfigure{\includegraphics[width=0.49\linewidth]{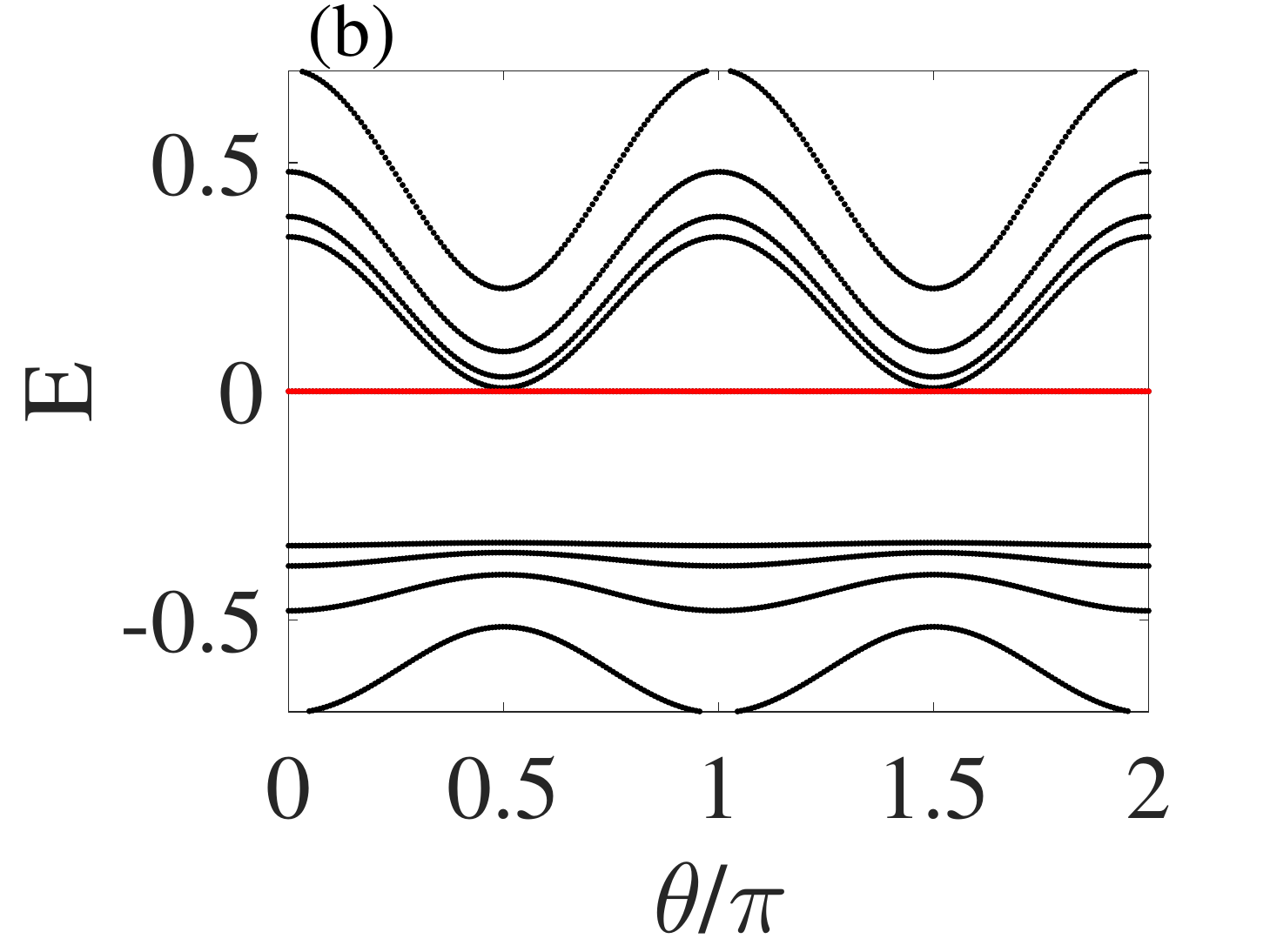}}
	
	\subfigure{\includegraphics[width=0.49\linewidth]{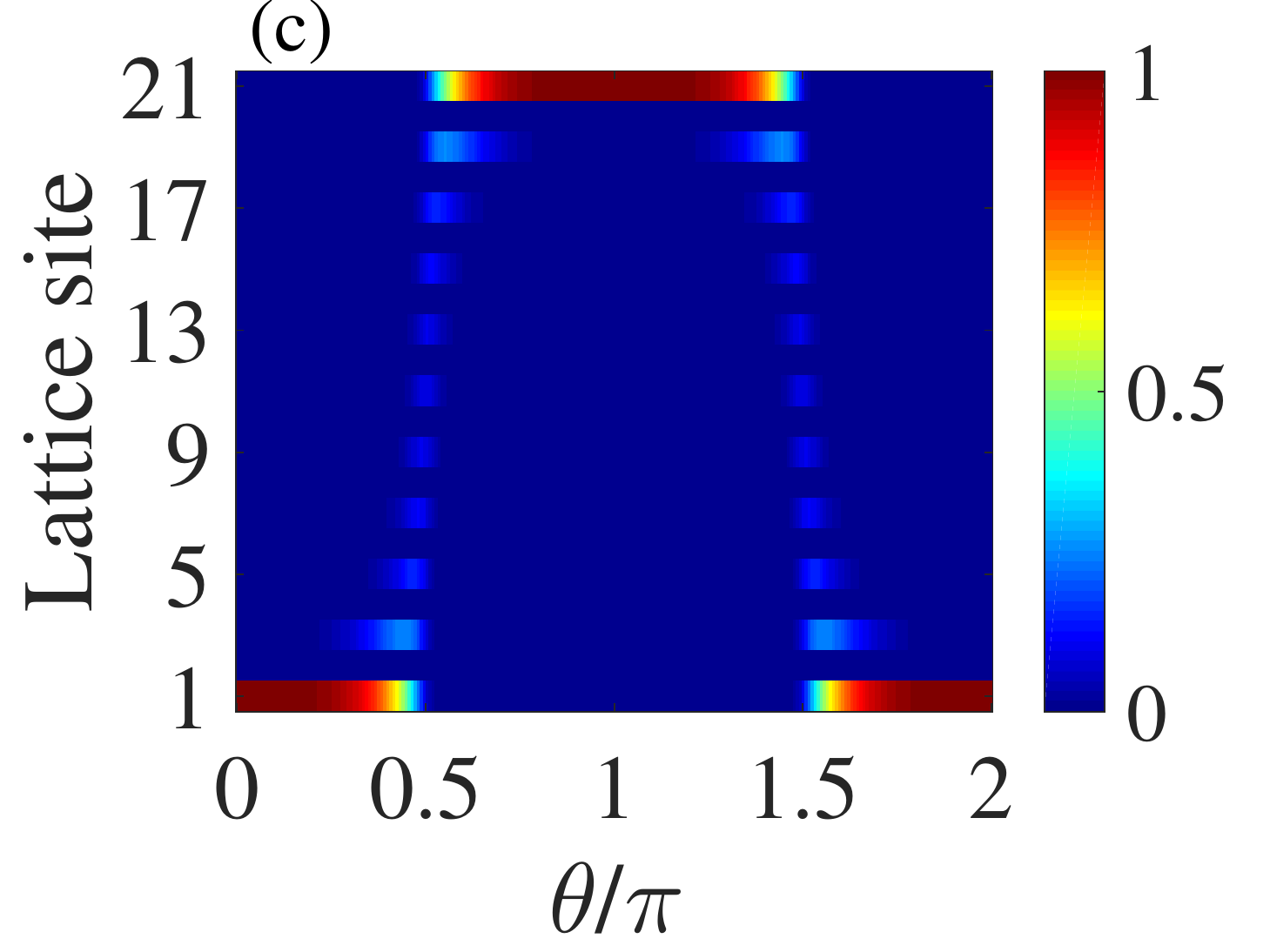}}
	\subfigure{\includegraphics[width=0.49\linewidth]{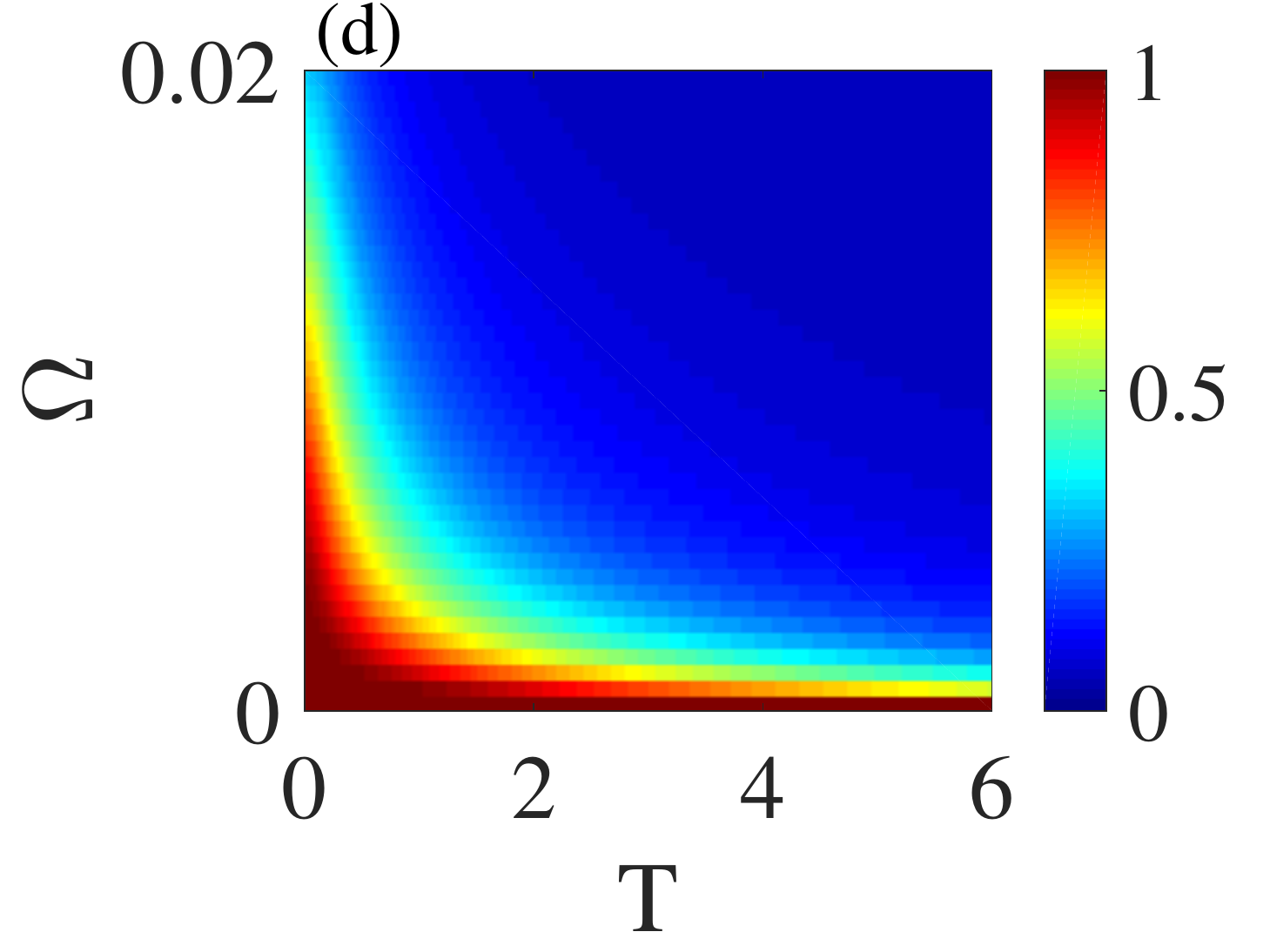}}
	\caption{(a) The energy spectrum of the system when $T=6$. (b) The detailed pattern in the energy gap. (c) The distribution of the gap state (red line in (b)). (d) The fidelity versus the varying rate $\Omega$ and partial NNN hopping strength $T$ when $T$ is only added on the even sites. The insert figure shows that there is a corresponding $\Omega$ to make the fidelity maximal when $T$ is large enough.}\label{fig6}
\end{figure} 

To further explore the reasons of the above results, we plot the energy spectrum and the corresponding distribution of the gap state when the partial NNN hopping strength $T$ is large enough, as shown in Fig.~\ref{fig4}. We find that the large enough $T$ makes the original zero energy mode become a wavy energy level, as shown in Figs.~\ref{fig4}(a) and \ref{fig4}(b). Although the wavy energy level is still located in the energy gap, its distribution becomes totally different, as shown in Fig.~\ref{fig4}(c). Obviously, the gap state is localized near the second site when $\theta\in (0, \frac{\pi}{2})\cup (\frac{3\pi}{2}, 2\pi)$ while it is localized near the penultimate site in other regions of $\theta$. It means that the initial state transfer channel between $|L\rangle$ and $|R\rangle$ becomes the state transfer channel between the phononic states of $|L^{'}\rangle=|0,~1,~0,~0,...,0,~0\rangle$ and $|R^{'}\rangle=|0,~0,~0,~0,...,1,~0\rangle$. To further demonstrate the above conclusions, we prepare the initial state in $|L^{'}\rangle$ and simulate the fidelity between the evolved final state and the state of $|R^{'}\rangle$ numerically, as shown in Fig.~\ref{fig5}(a). The numerical results show that the state transfer can be achieved with a high fidelity corresponding to a large enough $T$ and small enough $\Omega$. For example, when $T=6$ and $\Omega=0.00001$, the evolution of the initial phononic state $|L^{'}\rangle$ over time is shown in Fig.~\ref{fig5}(b). It means that we can achieve the state transfer between $|L^{'}\rangle$ and $|R^{'}\rangle$ via the gap state when $T$ is large enough. We stress that although $|L\rangle\rightarrow|R\rangle$ and $|L^{'}\rangle\rightarrow|R^{'}\rangle$ are different state transfer processes, both of them are achieved by the same topological channel (gap state). This means that we can realize the photonic and phononic state transfer processes only by one channel via designing the partial NNN hopping strength $T$ appropriately. Further, the two different kinds of state transfer can be switched to each other via modulating the value of $T$, such as opening the channel of $|L\rangle\rightarrow|R\rangle$ when $T<0.4$ and opening the channel of $|L^{'}\rangle\rightarrow|R^{'}\rangle$ when $T$ is large enough. Actually, we can also achieve the state transfer of  $|L\rangle \rightarrow |L^{'}\rangle$ ( $|R\rangle \rightarrow |R^{'}\rangle$) via varying $T$ over time for a given $\theta$, such as $H_{\mathrm{SSH}}(t)=\sum_{n} [(1-\cos\frac{\pi}{4}) a_{n}^{\dag}b_{n}+(1+\cos\frac{\pi}{4})a_{n+1}^{\dag}b_{n}+\Omega t a_{n+1}^{\dag}a_{n}]+\mathrm{H.c.}$. 

We also investigate the large enough partial NNN hopping added on the even sites (resonators), we find that the gap state still keeps the zero energy, as shown in Figs.~\ref{fig6}(a) and \ref{fig6}(b). The reason is that the NNN hopping added on the even sites has no effect on the two end sites, which makes the gap state keep the zero energy. It means that the state transfer channel between $|L\rangle$ and $|R\rangle$ keeps open and the large $T$ has no influence on the state transfer of $|L\rangle\rightarrow|R\rangle$, as shown in Fig.~\ref{fig6}(c). The fidelity versus the varying rate of $\theta$ and NNN hopping strength $T$ is shown in Fig.~\ref{fig6}(d). It can be seen clearly that there still exists a range of $\Omega$ to make the fidelity to be maximal corresponding to a large enough $T$. It means that we cannot achieve the different state transfers only by one gap state via designing $T$ added on the even sites.   

\section{\label{sec.3}Conclusions}
We have proposed a scheme to achieve the mapping of the SSH model with the partial NNN hopping based on a 1D frequency-modulated optomechanical array. We find that the quantum state transfer between the topological left edge state and the right edge state can be achieved via the topological channel located in the gap when the NNN hopping is vanishing. Meanwhile, the system can achieve another different quantum state transfer via the same topological channel when a large enough NNN hopping is added into the system. In this way, we can achieve two different kinds of quantum state transfer by one topological channel via controlling the NNN hopping strength. Our scheme provides a novel method to achieve different quantum state transfers in a controllable way.

\begin{acknowledgments}
This work was supported by the National Natural Science Foundation of China under Grant Nos. 61822114,  61575055, 11874132, 61465013, and 11465020, and the Project of Jilin Science and Technology Development for Leading Talent of Science and Technology Innovation in Middle and Young and Team Project under Grant No. 20160519022JH.
\end{acknowledgments}




\begin{thebibliography}{10}
	\newcommand{\enquote}[1]{``#1''}
	
	\bibitem{hasan2010colloquium}
	M.~Z. Hasan and C.~L. Kane, Colloquium: topological insulators, Rev. Mod. Phys.
	\textbf{82}, 3045 (2010).
	
	\bibitem{qi2011topological}
	X.-L. Qi and S.-C. Zhang, Topological insulators and superconductors, Rev. Mod. Phys. \textbf{83},
	1057 (2011).
	
	\bibitem{chiu2016classification}
	C.-K. Chiu, J.~C. Teo, A.~P. Schnyder, and S.~Ryu, Classification of topological quantum matter with symmetries, Rev.
			Mod. Phys. \textbf{88}, 035005 (2016).
	
	\bibitem{bansil2016colloquium}
	A.~Bansil, H.~Lin, and T.~Das, Colloquium: Topological band theory, Rev. Mod. Phys.
	\textbf{88}, 021004 (2016).
	
	\bibitem{matsuura2010momentum}
	S.~Matsuura and S.~Ryu, Momentum space metric, nonlocal operator, and topological insulators, Phys. Rev. B \textbf{82},
	245113 (2010).
	
	\bibitem{wray2011topological}
	L.~A. Wray, S.-Y. Xu, Y.~Xia, D.~Hsieh, A.~V. Fedorov, Y.~San~Hor, R.~J. Cava,
	A.~Bansil, H.~Lin, and M.~Z. Hasan, A topological insulator surface under strong Coulomb, magnetic and disorder perturbations, Nat. Phys.
	\textbf{7}, 32 (2011).
	
	\bibitem{nussinov2009sufficient}
	Z.~Nussinov and G.~Ortiz, Sufficient symmetry conditions for topological quantum order, Proceedings of the National
			Academy of Sciences \textbf{106}, 16944 (2009).
	
	\bibitem{shapourian2017many}
	H.~Shapourian, K.~Shiozaki, and S.~Ryu, Many-body topological invariants for fermionic symmetry-protected topological phases, Phys. Rev.
			Lett. \textbf{118}, 216402 (2017).
	
	\bibitem{dlaska2017robust}
	C.~Dlaska, B.~Vermersch, and P.~Zoller, Robust quantum state transfer via topologically protected edge channels in dipolar arrays, Quantum Science
			and Technology \textbf{2}, 015001 (2017).
	
	\bibitem{bonderson2011topological}
	P.~Bonderson and R.~M. Lutchyn, Topological quantum buses: coherent quantum information transfer between topological and conventional qubits, Phys. Rev. Lett.
	\textbf{106}, 130505 (2011).
	
	\bibitem{leijnse2011quantum}
	M.~Leijnse and K.~Flensberg, Quantum information transfer between topological and spin qubit systems, Phys. Rev. Lett.
	\textbf{107}, 210502 (2011).
	
	\bibitem{freedman2003topological}
	M.~Freedman, A.~Kitaev, M.~Larsen, and Z.~Wang, Topological quantum computation, Bulletin
			of the American Mathematical Society \textbf{40}, 31 (2003).
	
	\bibitem{sau2010generic}
	J.~D. Sau, R.~M. Lutchyn, S.~Tewari, and S.~D. Sarma,
	Generic new platform for topological quantum computation using semiconductor heterostructures, Phys. Rev. Lett. \textbf{104}, 040502 (2010).
	
	\bibitem{nayak2008non}
	C.~Nayak, S.~H. Simon, A.~Stern, M.~Freedman, and S.~D. Sarma,
	Non-Abelian anyons and topological quantum computation, Rev. Mod. Phys. \textbf{80}, 1083 (2008).
	
	\bibitem{sarma2015majorana}
	S.~D. Sarma, M.~Freedman, and C.~Nayak, Majorana zero modes and topological quantum computation, npj Quantum
			Inf. \textbf{1}, 15001 (2015).
	
	\bibitem{fradkin1983phase}
	E.~Fradkin and J.~E. Hirsch, Phase diagram of one-dimensional electron-phonon systems. I. The Su-Schrieffer-Heeger model, Phys. Rev. B \textbf{27},
	1680 (1983).
	
	\bibitem{marchand2010sharp}
	D.~Marchand, G.~De~Filippis, V.~Cataudella, M.~Berciu, N.~Nagaosa,
	N.~Prokof’ev, A.~Mishchenko, and P.~Stamp, Sharp transition for single polarons in the one-dimensional Su-Schrieffer-Heeger model, Phys.
			Rev. Lett. \textbf{105}, 266605 (2010).
	
	\bibitem{li2014topological}
	L.~Li, Z.~Xu, and S.~Chen, Topological phases of generalized Su-Schrieffer-Heeger models, Phys. Rev. B \textbf{89},
	085111 (2014).
	
	\bibitem{fu1988bound}
	R.~Fu, Z.~Shuai, J.~Liu, X.~Sun, and J.~C. Hicks, Bound states trapped by the soliton in the Su-Schrieffer-Heeger model, Phys. Rev. B \textbf{38}, 6298 (1988).
	
	\bibitem{dangel2018topological}
	F.~Dangel, M.~Wagner, H.~Cartarius, J.~Main, and G.~Wunner,
	Topological invariants in dissipative extensions of the Su-Schrieffer-Heeger model, Phys. Rev. A \textbf{98}, 013628 (2018).
	
	\bibitem{yao2018non}
	S.~Yao, F.~Song, and Z.~Wang, Non-hermitian chern bands, Phys. Rev. lett.
	\textbf{121}, 136802 (2018).
	
	\bibitem{zhu2014pt}
	B.~Zhu, R.~L{\"u}, and S.~Chen, PT symmetry in the non-Hermitian Su-Schrieffer-Heeger model with complex boundary potentials, Phys. Rev. A
	\textbf{89}, 062102 (2014).
	
	\bibitem{klett2018topological}
	M.~Klett, H.~Cartarius, D.~Dast, J.~Main, and G.~Wunner,
	Topological edge states in the Su-Schrieffer-Heeger model subject to balanced particle gain and loss, Eur. Phys. J. D \textbf{72}, 214 (2018).
	
	\bibitem{klett2017relation}
	M.~Klett, H.~Cartarius, D.~Dast, J.~Main, and G.~Wunner,
	Relation between PT-symmetry breaking and topologically nontrivial phases in the Su-Schrieffer-Heeger and Kitaev models, Phys. Rev. A \textbf{95}, 053626 (2017).
	
	\bibitem{ke2017topological}
	S.~Ke, B.~Wang, H.~Long, K.~Wang, and P.~Lu, Topological edge modes in non-Hermitian plasmonic waveguide arrays, Opt.
			express \textbf{25}, 11132 (2017).
	
	\bibitem{longhi2019probing}
	S.~Longhi, Probing topological phases in waveguide superlattices, Opt. Lett. \textbf{44}, 2530 (2019).
	
	\bibitem{tan2014photonic}
	W.~Tan, Y.~Sun, H.~Chen, and S.-Q. Shen, Photonic simulation of topological excitations in metamaterials, Sci. Rep.
	\textbf{4}, 3842 (2014).
	
	\bibitem{mei2015simulation}
	F.~Mei, J.-B. You, W.~Nie, R.~Fazio, S.-L. Zhu, and L.~C. Kwek,
	Simulation and detection of photonic Chern insulators in a one-dimensional circuit-QED lattice, Phys. Rev. A \textbf{92}, 041805 (2015).
	
	\bibitem{lieu2018topological}
	S.~Lieu, Topological phases in the non-Hermitian Su-Schrieffer-Heeger model, Phys. Rev. B \textbf{97}, 045106 (2018).
	
	\bibitem{kunst2018biorthogonal}
	F.~K. Kunst, E.~Edvardsson, J.~C. Budich, and E.~J. Bergholtz,
	Biorthogonal bulk-boundary correspondence in non-Hermitian systems, Phys. Rev. Lett. \textbf{121}, 026808 (2018).
	
	\bibitem{yao2018edge}
	S.~Yao and Z.~Wang, Edge states and topological invariants of non-Hermitian systems, Phys. Rev. Lett. \textbf{121},
	086803 (2018).
	
	\bibitem{mei2018robust}
	F.~Mei, G.~Chen, L.~Tian, S.-L. Zhu, and S.~Jia, Robust quantum state transfer via topological edge states in superconducting qubit chains, Phys. Rev. A \textbf{98}, 012331 (2018).
	
	\bibitem{singh2014optomechanical}
	V.~Singh, S.~Bosman, B.~Schneider, Y.~M. Blanter, A.~Castellanos-Gomez, and
	G.~Steele, Optomechanical coupling between a multilayer graphene mechanical resonator and a superconducting microwave cavity, Nature nanotechnology \textbf{9}, 820
	(2014).
	
	\bibitem{qi2017simulating}
	L.~Qi, Y.~Xing, H.-F. Wang, A.-D. Zhu, and S.~Zhang,
	Simulating $Z_{2}$ topological insulators via a one-dimensional cavity optomechanical cells array, Opt. Express \textbf{25}, 17948 (2017).
	
\end{thebibliography}

\end{document}